\documentclass[letterpaper,preprint]{revtex4}
\usepackage{graphicx}
\usepackage{subfigure}
\begin{document}
\date{\today}
\title{Implementation of a single femtosecond optical frequency comb for rovibrational cooling
}
\author{W. Shi, S. Malinovskaya\\
Department of Physics and Engineering Physics, Stevens Institute of
Technology, Hoboken, NJ 07030}

\begin{abstract}
We show that a single femtosecond optical frequency comb may be used
to induce two-photon transitions between molecular vibrational
levels to form ultracold molecules, e.g., KRb. The phase across an
individual pulse in the pulse train is sinusoidally modulated with a
carefully chosen modulation amplitude and frequency. Piecewise
adiabatic population transfer is fulfilled to the final state by
each pulse in the applied pulse train providing a controlled
population accumulation in the final state. Detuning the pulse train
carrier and modulation frequency from one-photon resonances changes
the time scale of molecular dynamics but leads to the same complete
population transfer to the ultracold state. A standard optical
frequency comb with no modulation is shown to induce similar
dynamics leading to rovibrational cooling.
\end{abstract}

\maketitle

An optical frequency comb has been recognized as 
a new and unique tool for high-resolution spectroscopic analysis of
internal energy structure and dynamics as well as for controlling
ultrafast phenomena in atomic and molecular physics
\cite{Ye05,St06,Th06,Di07}. Owing to its broadband spectrum, the
frequency comb may efficiently interact with the medium inducing
one-photon,
two-photon and multi-photon resonances between 
finely structured energy levels. A unique ability of the frequency
comb is provided by the presence of about a million of optical modes
in its spectrum with very narrow bandwidth and exact frequency
positions \cite{Os06}. During last years, the investigations have
been carried out on implemention of a femtosecond frequency comb to
manipulate ultracold gases. The pioneering works in quantum control
in ultracold temperatures include the two-photon excitation of
specific atomic levels forming four-level diamond configuration in
cold $^{87}Rb$ using a phase modulated, femtosecond optical
frequency comb, \cite{St08}, and a theory on piecewise stimulated
Raman adiabatic passage
 performed with two coherent
pulse trains possessing the pulse-to-pulse amplitude and chirped
phase variation, \cite{Pe07,Sh08}. In these papers, the authors
reported on creation of ultracold KRb molecules from Feshbach
states, the highest excited vibrational states of the ground
electronic state, using the pump-dump stepwise technique that
coherently accumulates population in the ground vibrational state.
Experimentally, a dense quantum gas of ultracold KRb polar molecules
was produced from the Feshbach molecules using the STIRAP scheme
with two $\mu s$ pulses, \cite{Ni08}. Coherent population transfer
was demonstrated to rovibrational ground state of the triplet and
singlet electronic ground potential.

In this paper,
 we demonstrate that a single, phase modulated optical frequency comb may be used to control
population dynamics aiming creation of deeply bound ultracold
molecules from Feshbach states. We investigate the KRb rovibrational
cooling, that involves the interaction of loosely bound KRb
molecules with a single femtosecond optical frequency comb resulting
in population flow from the Feshbach state to the ultracold state.
 The population dynamics takes place via
the two-photon Raman transitions that involve primary three energy
states separated in THz region. The efficient Raman transitions may
occur resonantly and also when the carrier frequency and the
modulation frequency of the laser field are detuned off resonance
with the one-photon transition frequency between
electronic-vibrational states. Coherent accumulation of population
in the cold KRb state with a negligible population of the excited
state is accomplished by a well defined number of sequential pulses
with zero carrier-envelope phase and within the lifetime of the
Feshbach KRb molecules, which is known to be about 100 ms,
\cite{Zi080,Zi08}.

A semi-classical model of two-photon Raman transitions, induced by a
femtosecond optical frequency comb in a three-level $\lambda$
system, describes the cooling process of the internal degrees of
freedom initiated in Feshbach molecules. Feshbach molecules can be
created by association of ultracold atom pairs performed by magnetic
field sweep across zero-energy resonance between the diatomic
vibrational bound state and the threshold for dissociation into an
atom pair at rest, \cite{Ko06}. They are the key intermediates in
the process of creation of deeply bound molecules in the ground
electronic configuration.

 A frequency comb known to be characterized
by two key parameters, the radio-frequency $f_r$, determined by the
pulse repetition rate and specifying the spacing between modes, and
by the carrier-envelop-offset frequency $f_0=f_r
\Delta\phi_{ce}/(2\pi)$, here $\Delta \phi_{ce}$ is the
carrier-envelope phase.
  Both parameters, the $f_0$ and the $f_r$, were
efficiently used to manipulate dynamics in, e.g., \cite{St08},
 to resonantly enhance two photon
transitions in cold $^{87}Rb$ atoms. Here, the control scheme is
simplified by not involving $f_0$. We make $f_0$ equal to zero by
implying zero carrier-envelope phase in each pulse in the pulse
train.

\begin{figure} \vspace{20pt}
\centerline{
\includegraphics[width=3in]{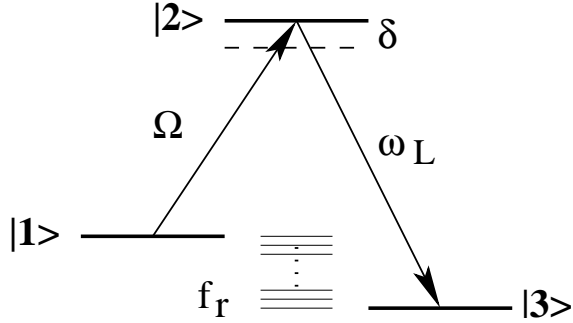}}
\caption{ The three-level $\lambda$-system modeling the molecular
energy levels, involved into cooling dynamics. State $|1>$ is the
Feshbach state, state $|2>$ is the transitional, electronic excited
state, and state $|3>$ is the cold molecular
state.}\label{lambda_system}
\end{figure}

The beam of a cw ring dye laser, having a sinusoidal modulation at
the MHz frequency, was implemented in \cite{Ha81} to study
absorption resonances in $I_2$ with high precision. Following the
idea of sin-phase modulation, we create a femtosecond pulse train,
having the sinusoidal modulation at THz frequency across an
individual pulse, aiming to induce two-photon Raman transitions. The
general form of the pulse train reads
\begin{equation}
\label{MFC}E(t,z)=\Sigma_{k=0}^{N-1} E_0 \exp(-(t-kT)^2/(2\tau^2))
\cos {(\omega_L (t-kT) + \Phi_0 \sin (\Omega (t-kT)))}.
\end{equation}
Here  k is an integer number and T is the period of the pulse train.
T is chosen to be much greater than a single pulse duration in order
for the wings of each pulse to reduce to zero before next pulse
arrives. The time-dependent phase across each pulse in the form of
the {\em sin} function enriches the frequency comb spectrum with new
peaks compared to a standard frequency comb. More specifically,
laser frequency $\omega_L$ determines the center of the frequency
comb, while sinusoidal modulation forms the sidebands at multiples
of $\Omega$ with the amplitude dictated by $\Phi_0$. The fine
structure of the optical broadband comb is owing to the modes
equally  spaced by the radio frequency $2\pi T^{-1}$. We apply a
single, sin-phase modulated optical frequency comb to induce Raman
resonances in the three-level $\lambda$-system aiming full
population transfer from the initial state $|1>$ through
intermediate state $|2>$ to the final state $|3>$,
Fig.(\ref{lambda_system}). We assume that state $|1>$ is the
Feshbach state, state $|2>$ is the transitional, electronic excited
state, or state manifold, and state $|3>$ is the ultracold molecular
state.  For Fechbach molecules, the energy splitting is large enough
to justify the validity of three-level model for the description of
their interaction with the frequency comb aiming to coherently
create ultracold molecules. 

 Parameters of the
$\lambda$-system correspond to data on molecular rovibrational
cooling of loosely bound KRb molecules from the Feshbach states,
discussed in \cite{Sh08}. Coherent population accumulation was
achieved in the deeply bound ground electronic state with
vibrational quantum number v=22. The related parameters of the
molecular system, implemented in our model, are $\omega_{21}$=340.7
THz, $\omega_{32}$=410.7 THz, and $\omega_{31}$=70 THz. In the
discussion below, all frequency and time parameters will be given in
the units of $[\omega_{31}]$ or $[\omega_{31}^{-1}]$ for convenience
of interpretation, then, e.g., $\omega_{21} = 4.9$, and
$\omega_{32}=5.9$.
 The field frequencies are
in resonance with the one-photon transitions in the three-level
$\lambda$-system, the carrier frequency $\omega_L=\omega_{32}$, and
the modulation frequency $\Omega=\omega_{21}$. The peak Rabi
frequency is $\Omega_R = \omega_{31}$, the pulse duration is
$\tau$=0.25, (3 fs), and the pulse train period is $T=6400 \tau$,
(20 ps), giving the radio frequency $ f_r=6.25 \cdot 10^{-4}$, (43
GHz). A single pulse area is estimated to be 0.2$\pi$.  The
modulation of the carrier frequency having value 410.7 THz assumed
to be done at frequency 340.7 THz. To achieve IR modulation, an
approach, described in \cite{Su08}, on efficient generation of a
Raman-type optical frequency comb in an enhancement cavity may be
applied. The technique provides with the whole comb bandwidth
covering 300-900 THz.

\begin{figure}
\vspace{20pt} \centerline{
\includegraphics[width=3in]{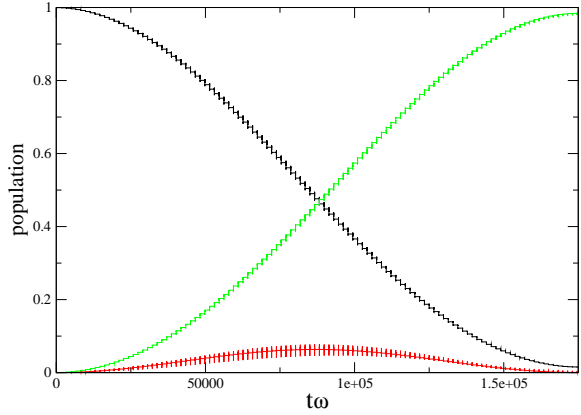}}
\caption{Population transfer in the three-level $\lambda$-system,
achieved via the resonant Raman transitions and using an optical
frequency comb as in Eq.(\ref{MFC}) having the carrier frequency and
the modulation frequency in resonance with one-photon transitions in
the $\lambda$-system. The values of the parameters are the carrier
frequency $\omega_L$=5.9, (410.7 THz), the modulation frequency
$\Omega$=4.9, (340.7 THz), the modulation amplitude $\Phi_0=4$, and
the peak Rabi frequency $\Omega_R$=1, (70THz); the system one-photon
transition frequencies are $\omega_{21}=\Omega$,
$\omega_{32}=\omega_L$. Stepwise, adiabatic accumulation of the
population is observed in state $|3>$, (green), which is the
ultracold KRb state. The population of the Feshbach state $|1>$,
(black), reduces gradually to zero, while the excited state manifold
$|2>$, (red), is slightly populated during the transitional time.
Full population transfer is accomplished in 112 pulses.
}\label{comb_cooling_AMP4_OM4.9_w_L5.9_frag}
\end{figure}

The evolution of the density matrix was investigated using the
 Leouville von Neuman equation with the interaction Hamiltonian
having two nonzero matrix elements $H_{ij}=\Omega_R(t-T) [\exp\{- i
((\omega_L + \omega_{ji})(t-T) + M(t-T))\} + \exp\{ i ((\omega_L -
\omega_{ji})(t-T) +  M(t-T))\}]$, here, i,j are the indexes of the
basis set, $i=1,2$ and $j=i+1$,
 $M(t-T)= \Phi_0 \sin \Omega (t-T)$ is the modulation of the phase, $\Omega_R(t-T)= \Omega_R
\exp{(-(t-T)^2/(2\tau^2))}$ is the Rabi frequency, and $\Omega_R$ is
the peak value of the Rabi frequency. Calculations were done beyond
the rotating wave approximation, that allows for flexible values of
the transition frequencies relative to the laser carrier frequencies
to be considered. To get zero value of the carrier-envelope phase,
the temporal variation is taken in the form of (t-T), where T is the
pulse train period. It guarantees the envelope maximum to coincide
with the peak value of the amplitude of the electric field. The
condition for the Raman resonance is satisfied
 by making the difference of laser field frequency components $(\omega_L - \Omega)$
 equal to
the two-photon transition frequency $\omega_{31}$. Additionally, the
modes that are multiples of the radio frequency $f_r$ provide with
pairs of optical frequencies that differ by exactly the transition
frequency $\omega_{31}$. These lead to an efficient stepwise
population accumulation in the final state in the $\lambda$-system.
 The results of population transfer are presented
in the Fig.(\ref{comb_cooling_AMP4_OM4.9_w_L5.9_frag}) for the
parameters: The peak Rabi frequency $\Omega_R$=1, the carrier
frequency $\omega_L$=5.9, the modulation frequency $\Omega$=4.9, the
pulse duration $\tau=0.25$, and the modulation amplitude $\Phi_0=4$.
The $\lambda$-system transition frequencies are $\omega_{21}$=4.9,
and $\omega_{32}$=5.9. Adiabatic population transfer is achieved
from the initial $|1>$ to the final $|3>$ state via the transitional
state $|2>$, which is insignificantly populated. A negligible
population of the intermediate state is favorable for minimizing
spontaneous emission losses. Each pulse brings a fraction of
population, ($\sim 1\%$), to the final state and contributes to the
accumulative effect. Total population transfer occurs after 112
sequential pulses and is accomplished in 2.5 ns, which is within the
lifetime of the Feshbach KRb molecules.

\begin{figure}
\vspace{20pt} \centerline{
\includegraphics[width=3in]{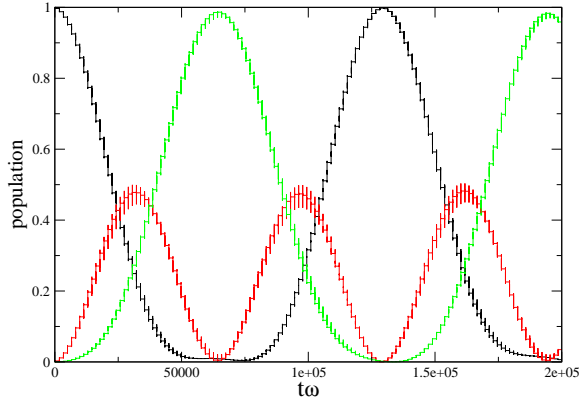}}
\caption{Population transfer in the three-level $\lambda$-system,
achieved using an optical frequency comb as in Eq.(\ref{MFC}) with
the field parameters detuned off resonance with the frequencies of
the $\lambda$-system. The detuning $\delta$ is equal to
$\omega_{31}/2$. The carrier frequency is $\omega_L$=5.4, the
modulation frequency is $\Omega$=4.4, the Rabi frequency is
$\Omega_R$=1, and the modulation amplitude is $\Phi_0=4$. The
$\lambda$-system transition frequencies are $\omega_{21}$=4.9,
$\omega_{32}$=5.9 (in the unites of frequency $\omega_{31}=70THz$).
Full population transfer to the final, cold state, (green), occurs
within 42 pulses. During this time, population of the initial
Feshbach state, (black), reduces to zero and the excited state,
(red), gets substantially populated during the transitional time.
The population is reversed by the next 42 pulses.
}\label{comb_cooling_AMP4_OM4.4_w_L5.4}
\end{figure}

When the carrier frequency $\omega_L$ and the modulation frequency
$\Omega$ of the optical frequency comb are detuned off resonance
with the transitional frequencies $\omega_{31}$ and $\omega_{21}$
respectively, the stepwise adiabatic population transfer takes place
in a fashion similar to the resonance case. The time evolution of
population
 is presented in the
Fig.(\ref{comb_cooling_AMP4_OM4.4_w_L5.4}) for parameters
$\Omega_R$=1, the carrier frequency $\omega_L$=5.4, the modulation
frequency $\Omega$=4.4, and the modulation amplitude $\Phi_0=4$; the
$\lambda$-system transition frequencies are $\omega_{21}$=4.9, and
$\omega_{32}$=5.9, giving $\delta=\omega_{31}/2$. Figure shows
 the coherent accumulation of the population in the final state,
taking place within 42 pulses and accomplished in about 1 ns.
However, the population of the transitional, electronic excited
state is substantial,
 almost $50\%$.
Compared to the resonance case, the population dynamics is faster
with less number of pulses in the pulse
train involved into full population transfer. 

\begin{figure}
\vspace{20pt} \centerline{
\includegraphics[width=3in]{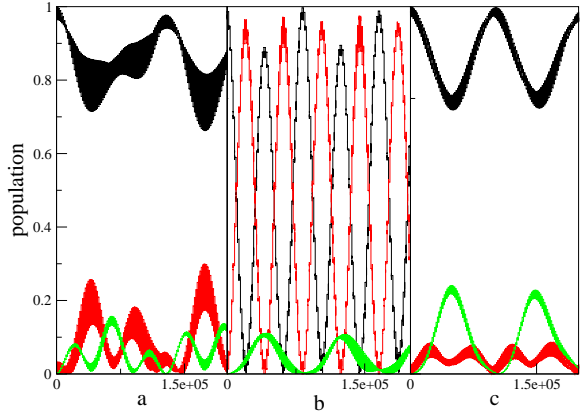}}
\caption{Population dynamics in three-level $\lambda$-system,
reached using an optical frequency comb as in Eq.(\ref{MFC}) with
different values of the modulation amplitude equal to
$\Phi_0$=3.(a), $\Phi_0$=5.(b), $\Phi_0$=8.(c). Other parameters are
the effective Rabi frequency $\Omega_R$=1, (70THz), the carrier
frequency $\omega_L$=5.9, (410.7 THz), and the modulation frequency
$\Omega$=4.9, (340.7 THz). For all three values of $\Phi_0$, there
is only partial population transfer to the final state $|3>$,
(green), unlike to the solution for $\Phi_0=4.$ in
Fig.(\ref{comb_cooling_AMP4_OM4.9_w_L5.9_frag}).
}\label{comb_cooling_AMPvario_resonance}
\end{figure}

\begin{figure}
\vspace{20pt} \centerline{
\includegraphics[width=3in]{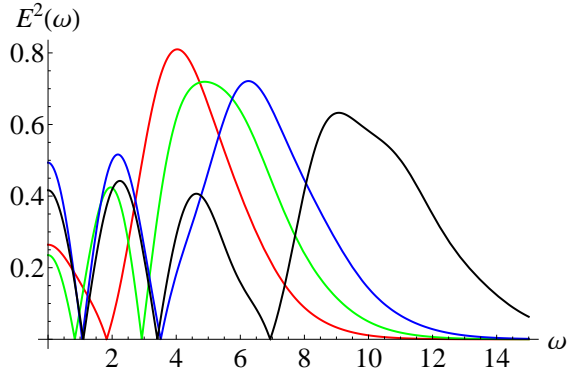}}
\caption{ The envelope of the power spectrum of the optical
frequency comb as in Eq.(\ref{MFC}), (no dense radio frequency comb
lines are included). The pulse train parameters are the effective
Rabi frequency $\Omega_R$=1, (70THz), the carrier frequency
$\omega_L$=5.9 ,(410.7 THz), the modulation frequency $\Omega$=4.9,
(340.7 THz), and the modulation amplitude $\Phi_0$=3. (red),
$\Phi_0$=4. (green), $\Phi_0$=5. (blue), $\Phi_0$=8.
(black).}\label{comb_cooling_FFT}
\end{figure}

Notably, there is a strong dependence of the efficiency of
population transfer on the value of the amplitude $\Phi_0$ of
sinusoidal modulation of the phase across individual pulse  in the
pulse train. For the resonant excitation, population dynamics was
calculated using different values of the parameter $\Phi_0$, the
results are presented in Fig.\ref{comb_cooling_AMPvario_resonance}
for $\Phi_0=3,5,8$. The desired adiabatic accumulation of the
population in the final state is not achieved for any of these
values of $\Phi_0$.
To understand the impact of the modulation amplitude $\Phi_0$, we
analyzed the Fourier transform of the pulse train in Eq.(\ref{MFC}).
The Fourier transform reads
\begin{equation}
E(\omega)=(E_0 \tau)/2 \Sigma_n J_n(\Phi_0) 
\exp{(-1/2(\omega_L+n\Omega-\omega)^2\tau^2)}
\cdot\Sigma_k \exp{(i \omega k T)}. \label{FFT_MOFC}
\end{equation}
Here, $J_n(\Phi_0)$ is the Bessel function of the order n and
$\Phi_0$ is the modulation index. When multiplied by
$\exp{(-1/2(\omega_L+n\Omega-\omega)^2\tau^2)}$, it determines the
shape of the power spectrum of the optical frequency comb. Depending
on the value of $\Phi_0$, the power spectrum has different number of
maxima as it is seen in Fig.{\ref{comb_cooling_FFT}} for
$\Phi_0=3,4,5,8$. The increase in modulation index brings
additional, intense peaks of modes into spectrum and broadens it.
These maxima are located at different frequencies for different
values of $\Phi_0$ affecting the population dynamics in the
$\lambda$-system.
The power spectrum of the pulse train with the modulation amplitude
$\Phi_0$=4 has three maxima, the highest one is at $\omega=4.9$
(which is in resonance with the $\omega_{21}$), making
 the pulse train with
$\Phi_0$=4, an optimal one for coherent accumulation of the
population in the final state of the $\lambda$-system. It provides
full population transfer to the ground electronic and rovibrational
state and, thus, cooling the internal degrees of freedom in the KRb
Feshbach molecule. Thus, the parameters of the phase stabilized
pulse train, needed to accomplish rovibrational cooling from
Feshbach states, have to be chosen based on the analysis of the
power spectrum of the sin-phase modulated optical frequency comb and
the energy levels involved into dynamics of the molecular system.

Lowering the value of the Rabi frequency, which corresponds to
decrease in the field intensity, gives qualitatively similar results
of coherent, adiabatic accumulation of population in the final,
ultracold state leading to full population transfer. The difference
is in the time scale of the dynamics which becomes an order of
magnitude longer, as calculations show, however, is well within the
lifetime of the Feshbach molecules. Elongation of duration of
population dynamics is observed also with the decrease in the pulse
repetition rate. Due to computational limitations on the propagation
time, we use the values of the parameters of the Rabi frequency and
pulse train repetition rate that demonstrate the phenomenon on a
shorter time scale. Close to experimental values parameters of the
field intensity and the repetition rate are implemented to the
standard optical frequency comb interaction with the three-level
$\lambda$-system, described next.

We investigated the interaction of a single, standard optical
frequency comb with the three-level $\lambda$-system. The standard
femtosecond optical frequency comb is formed by a phase-stabilized
pulse train without phase or amplitude modulation across an
individual pulse. The carrier-envelope phase is made to be zero. The
pulse train reads
\begin{equation}
\label{FC}E(t)=\Sigma_{k=0}^{N-1} E_0 \exp(-(t-kT)^2/(2\tau^2)) \cos
{(\omega_L (t-kT)))}.
\end{equation}
Here, T is the pulse train period, $\tau$ is the pulse duration,
$\omega_L$ is the carrier frequency. The carrier frequency of the
pulse train $\omega_L$ is chosen to be in resonance with the
one-photon transition frequency $\omega_{32}$ in the
$\lambda$-system. The two-photon resonances in the $\lambda$-system
are provided by the pairs of optical frequencies present within the
frequency comb that are multiples of the radio frequency and satisfy
the two-photon resonance condition $ m f_r - n f_r = \omega_L - n'
f_r =\omega_{31}$, here $m, n, n'$ are integer numbers. Calculations
were performed for two values of the peak Rabi frequency: $\Omega_R
= 0.1 \omega_{31}$ and $\Omega_R = 0.01 \omega_{31}$, they determine
the electric field intensity in the range from $ 10^{14}$ to $
10^{12}$ $W/cm^2$. A single pulse area is estimated to be 0.03$\pi$
to 0.003$\pi$ respectively. Two values of the pulse train period
were considered: $T=6.4\cdot10^4\tau$=0.2 ns, (5 GHz), and
$T=6.4\cdot10^5 \tau$=2 ns, (500 MHz), here the pulse duration
$\tau$ is 3 fs.

For the $\lambda$-system, we apply a set of parameters that
correspond to experimental data on molecular rovibrational cooling
of loosely
bound KRb molecules from the Feshbach states presented in 
\cite{Ni08}. In the described experiments, the STIRAP scheme was
implemented to coherently transfer population from Feshbach states
to the ground electronic triplet or the ground electronic singlet
state with zero rovibrational quantum number. These deeply bound
states were achieved
 with two $\mu s$ pulses that were
in resonance with the one-photon transitions between the molecular
states
 involved into the process, which are the
Feshbach state, the 2$^3\Sigma$ electronically excited state,
 and the triplet or singlet ground electronic states. In our model, we
address the fundamentally cold state of the KRb molecule by using
parameters that correspond to the experiment involving the singlet
ground electronic state.
 These parameters are the $\omega_{21}$=309.3
THz, and the $\omega_{32}$=434.8 THz, giving the frequency of the
two-photon transition $\omega_{31}$=125.5 THz.
Fig.{\ref{comb_cooling_Science_resonant_stnrd_T64000} shows the
population dynamics in the $\lambda$-system for $\Omega_R = 0.1
\omega_{31}$= 12.56 THz and T=0.2 ns, which results in full
population transfer to the ultracold state within 238 pulses during
47.6 ns.
\begin{figure}
\vspace{20pt} \centerline{
\includegraphics[width=3in]{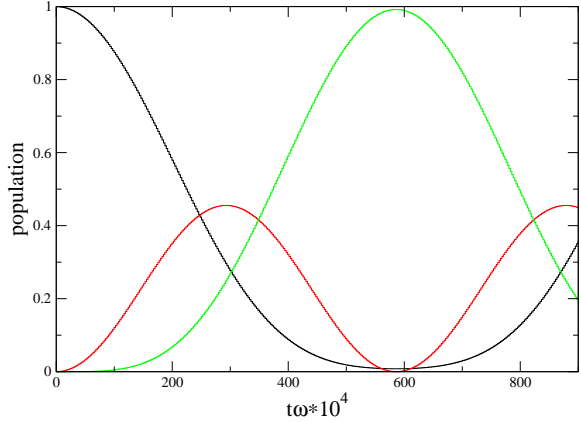}}
\caption{ Population transfer in the three-level $\lambda$-system
using a standard optical frequency comb characterized by the
radio-frequency equal to 5 GHz and zero offset frequency. Black
curve shows population dynamics of the Feshbach state, red - same
for the excited electronic state, green - same for the final,
ultracold state. Parameters of the pulse train are the carrier
frequency $\omega_L=\omega_{32}$=434.8 THz, and the pulse duration
$\tau_0$=3 fs.}\label{comb_cooling_Science_resonant_stnrd_T64000}
\end{figure}
Next 238 pulses restore full population in the initial state $|1>$ -
 the Feshbach state - and the dynamics repeats. Within the lifetime
of the Feshbach state, it is possible to transform the medium from
highly vibrationally excited molecules to the ultracold
configuration and back using a single, phase-locked pulse train.
Note, that qualitatively same result was obtained for the pule train
repetition rate 500 MTz, ($T=2 ns$), with the only difference in the
dynamics duration, which was extended by an order of magnitude to
440 ns.

Off-resonant Raman transitions are readily induced by a single,
standard optical frequency comb, provided there are optical
frequencies
 in the comb that satisfy the two-photon resonance condition and differ by exactly
 the
Feshbach-to-ultracold state transitional frequency. In
Fig.{\ref{comb_cooling_wL_2.0_w32_2.46_w31_1_AIN0.1_and_AIN_0.01_stadrd}
the case for the carrier frequency detuned off resonance with the
$\omega_{32}$ transitional frequency is presented for detuning
$\delta=3/2 \omega_{31}$ and two values of the Rabi frequency, the
$\Omega_R = 0.1 \omega_{31} =12.55 THz$ and $\Omega_R = 0.01
\omega_{31} = 1.26 THz$ related to the field intensity on the order
of $10^{14}$ and $10^{12}$ $W/cm^2$ respectively. The chosen value
of detuning $\delta$ is about $\sim 10^{14} Hz$. Bold black, bold
red, and bold green lines show population dynamics in the Feshbach
state, the excited electronic state and the final, ultracold state
respectively,
 for the Rabi frequency $\Omega_R=1.26 THz$, thin
lines  - same for the Rabi frequency $\Omega_R=12.55 THz$. Since
here we demonstrate the effects of the detuning, and the impact of
different field intensity on the population dynamics to the deeply
bound rovibrational state in the KRb molecule, the pulse train
period is chosen to be short to accommodate the computational
limitations, $T = 6400\tau = 0.02$ ns. (Same as for the resonance
case, pulse train period in 2 ns elongates the duration of
population accumulation.)
\begin{figure}
\vspace{20pt} \centerline{
\includegraphics[width=3in]{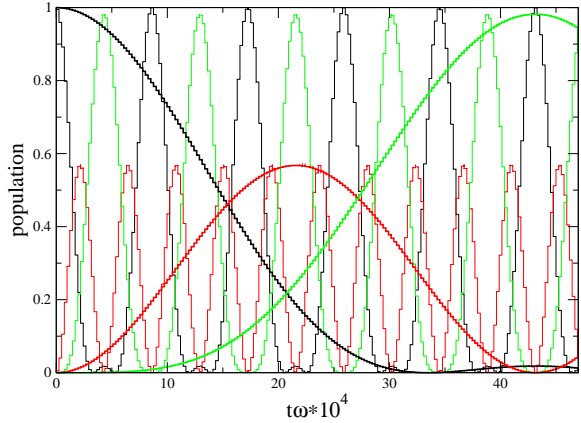}}
\caption{Population transfer in the three-level $\lambda$-system,
achieved using an optical frequency comb as in Eq.(\ref{FC}) with
the carrier frequency of the electric field detuned off resonance
with the $\omega_{32}$ in the $\lambda$-system. The detuning
$\delta$ is $ 3/2 \omega_{31}$. Bold black, bold red, and bold green
lines show population of the Feshbach state, the excited electronic
state and the final, ultracold state for the $\Omega_R$=1.26 THz,
($~10^{12}$  $W/cm^2$), thin lines show respective populations for
$\Omega_R$=12.55 THz, ($~10^{14}$ $W/cm^2$). Pulse train period is
$T = 6400\tau = 0.02 ns$.
}\label{comb_cooling_wL_2.0_w32_2.46_w31_1_AIN0.1_and_AIN_0.01_stadrd}
\end{figure}
Figure
Fig.{\ref{comb_cooling_wL_2.0_w32_2.46_w31_1_AIN0.1_and_AIN_0.01_stadrd}
shows that almost complete population transfer, $(\sim 98\%)$, takes
place when the carrier frequency is off-resonance with the
one-photon transitions in the molecular system. It also demonstrates
the field intensity dependence of the duration needed to accumulate
population in the ultracold state.  For the $\Omega_R$=12.55 THz,
($~10^{14}$ $W/cm^2$), the population transfer dynamics is
accomplished five times faster than for the $\Omega_R$=1.26 THz,
($~10^{12}$  $W/cm^2$).

In conclusion, we have demonstrated a possibility of molecular
rovibrational cooling from the Freshbach states using a single
femtosecond, optical frequency comb. We focused on the KRb molecule
and modeled it by a three-level $\lambda$-system with the energy
levels taken from \cite{Sh08} and \cite{Ni08}. Coherent accumulation
of the population in the cold state is achieved by applying a
standard optical frequency comb with zero offset frequency, and by
an optical frequency comb generated by a pulse train with a phase
modulation in the form of the sinusoidal function across an
individual pulse. An optimal dynamics takes place when the
carrier-envelope phase is equal to zero, ($f_0=0$). The mechanism of
the accumulative effect leading to full population transfer is based
on the excitation of the two-photon Raman resonances. For both, the
resonant and detuned Raman transition to happen, the radio frequency
$f_r$ characterizing the optical frequency comb has to be such that,
 multiplied by
 integers n and m, it forms the modes at the pump and Stokes
frequencies, whose difference, $n f_r - m f_r$, is in resonance with
the two-photon Raman transition frequency. In the case of
sinusoidally modulated optical frequency comb, the Raman transitions
are stimulated by the carrier and the modulation frequencies. A
special attention has to be paid to adjusting the modulation
amplitude (the index of the Bessel function, describing the optical
frequency comb), based on the comparison of the power spectrum with
the energy levels in a molecular system to be cooled. Compared to
\cite{Pe07,Sh08}, implementation of a single optical frequency comb
rather than two combs is more robust for experimental realization. A
single femtosecond optical frequency comb, with no modulation or
sinusoidally phase modulated, induces coherent adiabatic
accumulation of population in the ground state. Similar to the
mechanism behind the two frequency combs approach, a small fraction
of population is transferred by each pulse in the pulse train due to
its very small pulse area. When sinusoidal phase modulation is
applied across individual pulse, it results in negligible population
of the intermediate, electronically excited state which is a
valuable result in light of preserving population and minimizing the
spontaneous emission losses.

S.M. acknowledges helpful discussions with  J. Ye, P. Berman, and V.
Malinovsky. This research is supported by the National Science
Foundation under Grant No. PHY-0855391.

{}

\enddocument